\documentclass[aps,prl]{revtex4}

\usepackage{amsmath}
\usepackage{amssymb}
\usepackage{graphicx}
\usepackage{bm}
\usepackage{dcolumn}
\usepackage{color}
\usepackage{siunitx}
\begin{document}

\title{The evolution of electron dispersion in the series of rare-earth tritelluride compounds obtained from their charge-density-wave properties and susceptibility calculations}

\author{Pavel A. Vorobyev $^{1}$, Pavel D. Grigoriev $^{2,3,4}$, Kaushal K. Kesharpu $^{3}$ and Vladimir V. Khovaylo $^{3,5}$}

\address{$^{1}$ M.V. Lomonosov Moscow State University, Moscow 119991, Russia}
\email{pavel-vorobyev@mail.ru}

\address{$^{2}$ L.D. Landau Institute for Theoretical Physics, Chernogolovka 142432, Russia}

\address{$^{3}$ National University of Science and Technology MISiS, Moscow 119049, Russia}

\address{$^{4}$ P.N. Lebedev Physical Institute of RAS, Moscow 119991, Russia}

\address{$^{5}$ National Research South Ural State University, Chelyabinsk 454080, Russia}

\begin{abstract}
	We calculated electron susceptibility of rare-earth tritelluride compounds RTe$_3$ as a function of temperature, wave vector and electron-dispersion parameters. Comparison of results obtained with the available experimental data on the transition temperature and on the wave vector of a charge-density wave in these compounds allowed us to predict values and the evolution of electron-dispersion parameters with the variation of atomic number of rare-earth element R.
\end{abstract}

\maketitle


\section{Introduction}
In the last two decades the rare-earth tritelluride compounds RTe$_{3}$
(R=rare-earth elements) were actively studied, both theoretically\cite%
{Yao06,Hong2006} and experimentally by various techniques\cite%
{DiMasi95,Brouet2008,Moore2010,Schmitt2011,NRu2008,Ru2008,Zocco2015,Lavagnini10R,Fang07,BB,Hu2014,Kogar2019,Sinchenko2014,SinchPRB2012,SinchPRB2017,SinchJETPL2018}. A very rich electronic phase
diagram and the interplay between different types of electron ordering,\cite%
{NRu2008,Ru2008,Zocco2015} as well as amazing physical effects
in electron transport even at room temperature\cite%
{Sinchenko2014,SinchPRB2012,SinchPRB2017} stimulate this interest. These
compounds undergo a transition to a unidirectional charge-density wave (CDW)
state with wave vector $\boldsymbol{Q}_{CDW1}\approx (0,0,2/7c^{\ast })$. The corresponding
transition temperature $T_{CDW1}$ decreases with the atomic number of
rare-earth element R:\cite{NRu2008} $T_{CDW1}$ drops from over $600$K in LaTe%
$_{3}$ \cite{Kogar2019} to $T_{CDW1}=244$K in TmTe$_{3}$. However, the CDW
energy gap does not completely cover the Fermi surface (FS), as can be seen
from the ARPES measurements \cite{Brouet2008,Moore2010,Schmitt2011},
and the electronic properties below $T_{CDW1}$ remain metallic with a
reduced density of electron states at the Fermi level. In RTe$_{3}$
compounds with heaviest rare-earth elements the second CDW emerges\cite%
{NRu2008} with the wave vector $\boldsymbol{Q}_{CDW2}\approx (2/7a^{\ast },0,0)$ and the
transition temperature $T_{CDW2}$ increasing with the atomic number of the
rare-earth element R \cite{NRu2008} from $T_{CDW2}=52$K in DyTe$_{3}$ to $%
T_{CDW2}=180$K in TmTe$_{3}$. After the second CDW the RTe$_{3}$ compounds remain
metallic, similarly to NbSe$_{3}$. A third CDW has been proposed\cite{Hu2014}
from the optical conductivity measurements, but not yet confirmed by the
X-ray studies. At lower temperature the RTe$_{3}$ compounds become
magnetically ordered.\cite{Ru2008} In addition to all this, at high pressure the
RTe$_{3}$ compounds become superconducting.\cite{Zocco2015} 

To understand the richness of this phase diagram and the physical properties
in each phase it is very helpful to have information about the evolution of
electronic structure of RTe$_{3}$ compounds with the change of the atomic
number of rare-earth element R. Unfortunately, the ARPES data are available
only for very few compounds of this family and, in spite of a notable
progress in instrumentation, still have a large errorbar. The electron transport measurements are much more sensitive but they only give indirect information about the electronic structure, because of
a large number of electron scattering mechanisms.\cite%
{Sinchenko2014,SinchPRB2012,SinchPRB2017} In this paper we use the extensive
experimental data on the evolution of the CDW$_{1}$ wave vector $Q_{CDW1}$
and transition temperature $T_{c}$ to study the evolution of the electronic
structure of $RTe_{3}$ compounds. We calculate the electron susceptibility,
responsible for CDW$_{1}$ instability, as a function of the wave vector and
temperature at various parameters, which determine the electron dispersion.
The comparison of the results obtained with available experimental data
allows us making predictions about the evolution of these electron-structure
parameters with the atomic number of rare-earth element R.
 
\section{Calculation}

At temperatures $T>T_{CDW1}$ the in-plane electron dispersion in RTe$_{3}$
is described by a 2D tight binding model of the Te plane as developed in 
\cite{Brouet2008}, where the square net of Te atoms in each conducting layer 
forms two orthogonal chains created by the in-plane $p_x$ and $p_z$ orbitals. 
Correspondingly, $x$ and $z$ are the in-plane directions. 
In this model $t_{\parallel}$ and $t_{\perp}$ are the hopping amplitudes 
(transfer integrals) parallel and perpendicular to the 
direction of the considered $p$ orbital. The resulting in-plane electron 
dispersion can be written down as
\begin{equation}
\begin{split}
\varepsilon _{1}\left( k_{x},k_{z}\right) =& -2t_{\parallel }\cos \left[
\left( k_{x}+k_{z}\right) a/2\right] 
 -2t_{\perp }\cos \left[ \left( k_{x}-k_{z}\right) a/2\right] -E_{F}, \\
\varepsilon _{2}\left( k_{x},k_{z}\right) =& -2t_{\parallel }\cos \left[
\left( k_{x}-k_{z}\right) a/2\right] 
 -2t_{\perp }\cos \left[ \left( k_{x}+k_{z}\right) a/2\right] -E_{F},
\end{split}
\label{Disp}
\end{equation}%
where the calculated parameters for DyTe$_{3}$ are $t_{\parallel}=1.85$ eV, $%
t_{\perp}=0.35$ eV \cite{Brouet2008} and the in-plane lattice constant 
$a\approx 4.305\si{\angstrom} $ \cite{Ru2008}. The Fermi energy $E_F$ is determined from 
the electron density, namely, from the condition of 1.25 electrons for each $p_{x}$ and $p_{z}$ orbitals.\cite{Brouet2008} This condition gives us $E_{F}=-2t_{\parallel}\cos(%
\pi(1-\sqrt{3/8}))$. It is slightly (by 10\%) less than the originally used Fermi-energy value
$E_{F}=-2t_{\parallel}\sin(\pi/8)$, inaccurately determined \cite{Brouet2008} from the same condition (see Appendix).  
The resulting expression shows
the relation between these two parameters $t_{\parallel}$ and $E_{F}$, which is
important because they both affect the electron susceptibility.

For calculation we use the Kubo formula for the susceptibility of quantity 
$A$ with respect to quantity $B$ (see \S 126 of \cite{Landau}): 
\begin{equation}
\chi \left( \omega \right) =\frac{i}{\hbar }\int_{0}^{\infty }\left\langle %
\left[ \hat{A}\left( t\right) ,\hat{B}\left( 0\right) \right] \right\rangle
e^{i\omega t}dt.  \label{chi0}
\end{equation}%
For the free electron gas in the terms of matrix elements it becomes 
\begin{equation}
\chi \left( \omega \right) =\sum_{ml}A_{ml}B_{lm}\frac{n_{F}\left(
	E_{m}\right) -n_{F}\left( E_{l}\right) }{E_{l}-E_{m}-\omega -i\delta },
\label{chi1}
\end{equation}%
where $m$ and $l$ denote the quantum numbers $\left\{ \boldsymbol{k},s,\alpha
\right\} $, which are the electron momentum $\boldsymbol{k}$, spin $s$, and
the electron band index $\alpha $. In the CDW response function the quantities $%
A$ and $B$ are the electron density, so that Eq. (\ref{chi0}) is a
density-density correlator. To study the CDW onset one needs the 
static susceptibility at $\omega =0 $ but at a finite wave vector $\boldsymbol{Q}$. 
Electron spin only leads to a factor $4$ in susceptibility, but the summation over band index $\alpha $
must be retained if there are more than one band crossing the Fermi level. 
As a result we have for the real part of electron susceptibility 
\begin{equation}
\chi \left( \boldsymbol{Q}\right) =\sum_{\alpha ,\alpha ^{\prime }}\int 
\frac{4d^{d}\boldsymbol{k}}{\left( 2\pi \right) ^{d}}\frac{n_{F}\left( E_{ 
		\boldsymbol{k},\alpha }\right) -n_{F}\left( E_{\boldsymbol{k+Q},\alpha
		^{\prime }}\right) }{E_{\boldsymbol{k+Q},\alpha ^{\prime }}-E_{\boldsymbol{k}
		,\alpha }},  \label{chi2}
\end{equation}
where $n_{F}(\varepsilon )=1/\left( 1+\exp \left[ (\varepsilon -E_{F})/T%
\right] \right) $ is the Fermi-Dirac distribution function, d is the
dimension of space. Since the dispersion in the interlayer $y$-direction 
is very weak in RTe$_3$ compounds, we can take $d=2$. 
And each of the band indices $\alpha $ and $\alpha ^{\prime }$ may take 
any of two values $1,2$, because in RTe$_3$ two electron
bands cross the Fermi level. Here we assume that the matrix elements 
$A_{ml}$ and $B_{lm}$ do not depend on the band index. This means 
that due to the e-e interaction the electrons may scatter to any of 
the two bands with equal amplitudes. This assumption has virtually no effect on both temperature and $\boldsymbol{Q}$-vector dependence of the electron susceptibility, 
because the latter is determined mainly by the diagonal 
(in the band index) terms, which are enhanced in RTe$_3$ by a good nesting.    

Using Eq. (\ref{chi2}) we calculate the electron susceptibility $\chi$ as a function of 
CDW wave vector $\boldsymbol{Q}$ and temperature for various parameters 
$t_{\parallel }$ and $t_{\perp }$ of the bare electron dispersion (\ref{Disp}). 
The CDW phase transition happens when $\chi \left( \boldsymbol{Q},T\right) U=1$, where the interaction 
constant $U$ only weakly depends on the rare-earth atom R in RTe$_3$ family. 
The position of susceptibility maximum $\chi \left( \boldsymbol{Q}\right) $ 
gives the wave vector $\boldsymbol{Q}_{CDW1}$ of CDW instability as a function of 
the band-structure parameters $t_{\parallel }$ and $t_{\perp }$. 
The value of susceptibility in its maximum as a function of temperature 
$\chi_{max} \left( T \right) $ gives the evolution of CDW transition temperature 
$T_{CDW1}$ as a function of  $t_{\parallel }$ and $t_{\perp }$.

\section{Results and Discussion}

\begin{figure}[tbh]
	\centering
	\includegraphics[width=0.78\linewidth]{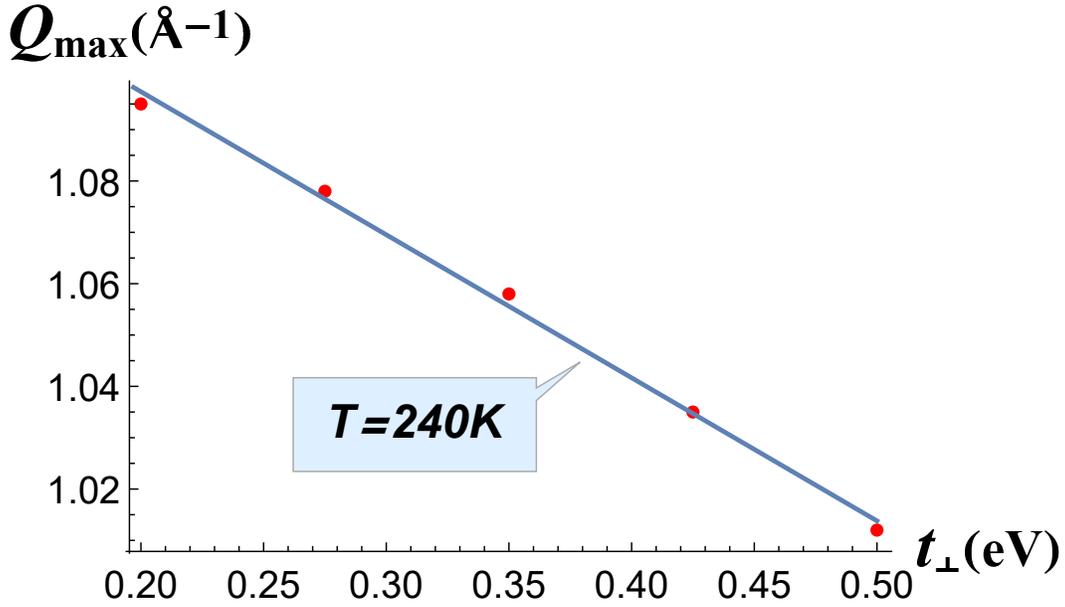} 
	\caption{CDW$_{1}$ wave vector $Q_{max}$ calculated at T=240 K as a function of the electron hopping term $t_{\perp}$.}
	\label{FigChitPerp1} 
\end{figure}

First we analyze the evolution of CDW$_1$ wave vector. The experimentally observed dependence of $%
Q_{CDW1}$ on the atomic number of rare-earth atom R can be taken, e.g., from Ref. \cite{Kogar2019}: 
$Q_{CDW1}$ monotonically increases by $\approx 10$\% with the increase of R-atom number from 
$Q_{CDW1}\approx 0.275$ r.l.u. in LaTe$_3$ to $Q_{CDW1}\approx 0.303$ r.l.u. in TmTe$_3$. 
The dependence of CDW wave vector $c$-component, $\boldsymbol{Q}_{CDW1}=(0,0,Q_{CDW1})$, 
on the perpendicular hopping term $t_{\perp}$, 
calculated using Eq. (\ref{chi2}), is shown in Fig.\ref{FigChitPerp1}. 
As we can see from this graph, $Q_{CDW1}(t_{\perp})$ demonstrates approximately linear dependence. 
The value $t_{\perp}$=0.35 eV, proposed in Ref. \cite{Brouet2008} from the band structure calculations, 
is located in the middle of this plot. The obtained $Q_{CDW1}(t_{\perp})$ dependence is rather weak: 
while $t_{\perp}$ increase dramatically, from $0.2$ to $0.5$eV, $Q_{CDW1}$ changes 
by only $\sim 8\% $ in $\si{\angstrom} ^{-1}$. In the reciprocal 
lattice units (r.l.u.) this variation is slightly stronger, as the lattice constant $c$ decreases 
with the atomic number from $c\approx 4.407\si{\angstrom} $ in LaTe$_3$ to $c\approx 4.28 \si{\angstrom} $ in ErTe$_3$ and TmTe$_3$, 
and the r.l.u. correspondingly increases in $\si{\angstrom} ^{-1}$. 
However, just the $Q_{CDW1}(t_{\perp})$ dependence cannot explain the observed evolution 
of the CDW$_1$ wave vector with the R-atom number, because it is too weak.

\begin{figure}[bbh]
	\centering
	\includegraphics[width=0.78\linewidth]{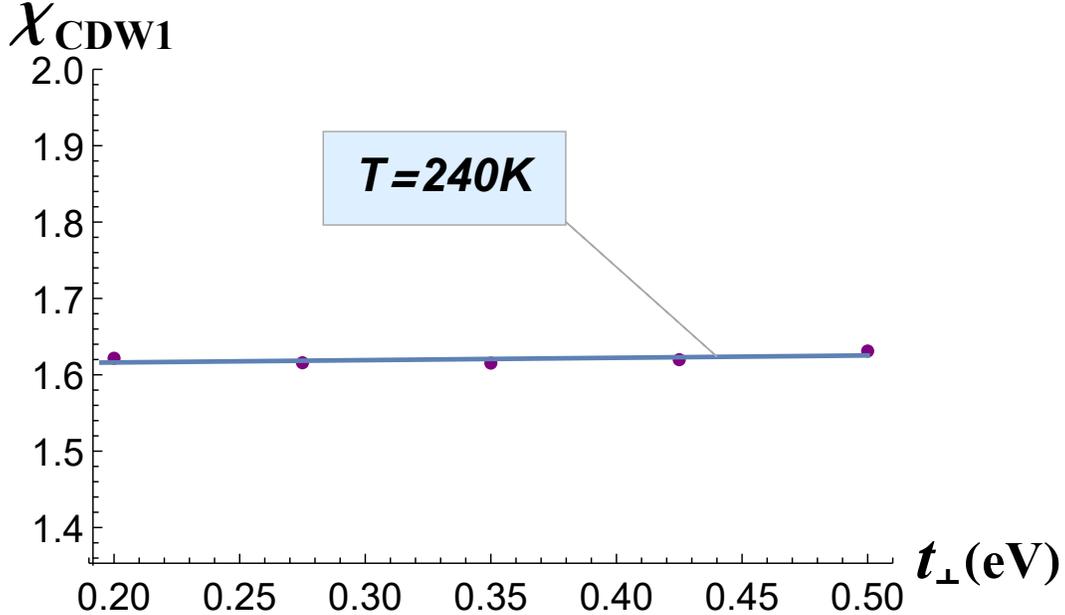} 
	\caption{Electron susceptibility $\protect\chi$ calculated at T=240 K
		as a function of the electron hopping term $t_{\perp}$.}
	\label{FigChitPerp2} 
\end{figure}

The dependence $\chi$($t_{\perp}$) is shown in Fig.\ref{FigChitPerp2}. The electron
susceptibility varies within one percent of its' maximum value and thus
remains almost constant. The $\chi_{CDW1}$ values are calculated on the wave vectors $%
Q_{CDW1}$, obtained for each value of $t_{\perp}$ as a position of the susceptibility maximum. 
From this plot we conclude that the parameter $t_{\perp}$ has almost no effect on the CDW$_1$
transition temperature. Hence, to interpret the evolution of CDW$_1$ transition temperature  $T_{CDW1}$
and of its wave vector $Q_{CDW1}$ with the rare-earth atomic number, 
one needs to consider their $t_{\parallel}$-dependence.

\begin{figure}[tbh]
	\centering
	\includegraphics[width=0.78\linewidth]{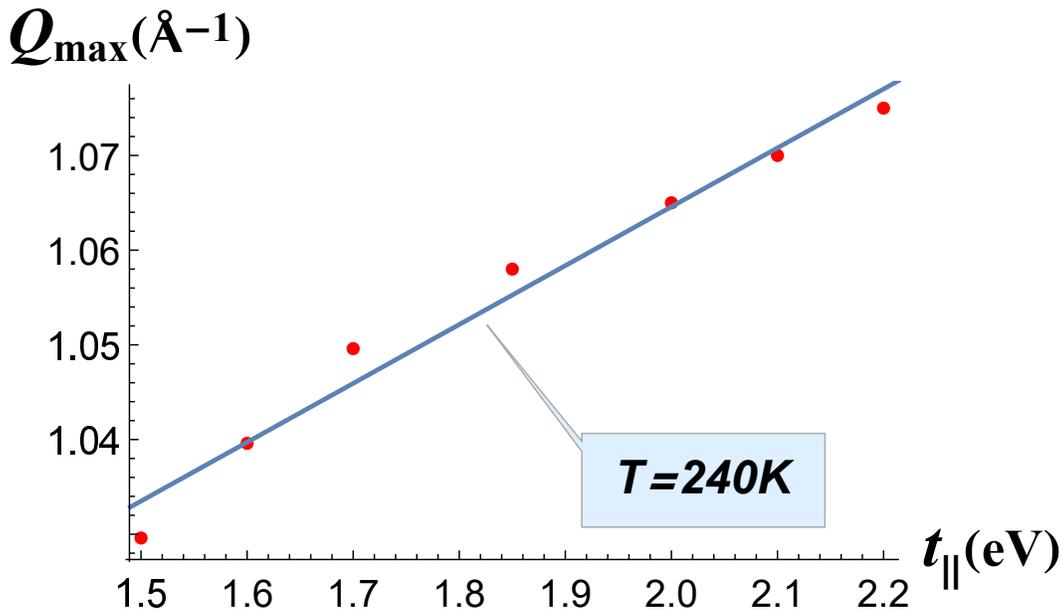} 
	\caption{CDW$_{1}$ wave vector $Q_{max}$ calculated at T=240 K as a function of the electron hopping term $t_{ \| }$.} \label{FigChitPar1} 
\end{figure}

The dependence $Q_{CDW1}(t_{\parallel})$ is shown in Fig. \ref{FigChitPar1}. The interval of this plot comprises 
the values $t_{\parallel}=1.7 eV$ and $t_{\parallel}=1.9$ eV, obtained in Ref. \cite{Brouet2008} 
from the band structure calculations for the lightest and heaviest rare-earth elements. 
$Q_{CDW1}$($t_{\parallel}$) demonstrates sublinear monotonic dependence, but $Q_{CDW1}$ there
increases with the increasing of parameter $t_{\parallel}$. 
It is opposite to the dependence $Q_{CDW1}(t_{\perp})$. 
Comparing Fig. \ref{FigChitPar1} with the experimental data on $Q_{CDW1}$, summarized in Ref. \cite{Kogar2019},
we may conclude that the parameter $t_{\parallel}$ increases with the atomic
number of the rare-earth element. According to the band structure calculations in Ref. \cite{Brouet2008} 
this transfer integral indeed increases from $t_{\parallel}=1.7 eV$ in LaTe$_{3}$ 
to $t_{\parallel}=1.9$ eV in LuTe$_{3}$. Thus, our conclusion 
qualitatively agrees with the band-structure calculations in Ref. \cite{Brouet2008}. 
However, according to our calculation the variation of $t_{\parallel}$ with the atomic
number of the rare-earth element must be stronger in order to account for the observed $Q_{CDW1}$ dependence.

\begin{figure}[tbh]
	\centering
	\includegraphics[width=0.78\linewidth]{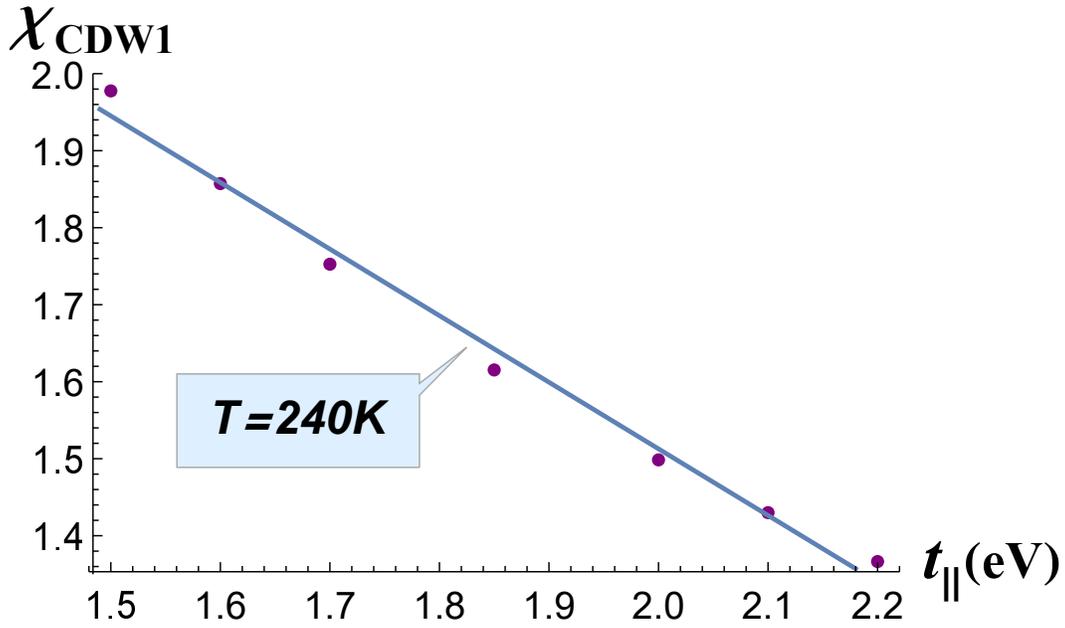} 
	\caption{Electron susceptibility $\protect\chi$ calculated at T=240 K
		as a function of the electron hopping term $t_{\|}$.} 
	\label{FigChitPar2} 
\end{figure}

In Fig. \ref{FigChitPar2} we plot the calculated $\chi (t_{\parallel})$ dependence, 
which is approximately linear. Similarly to our calculations of $\chi (t_{\perp})$, 
the susceptibility value is taken in its maximum as a function of the wave vector $Q_{CDW1}$. 
$\chi$
changes significantly -- about 35\% of its maximum value in the full range of
parameter $t_{\parallel}$ change. The $CDW_{1}$ transition temperature $T_{c}$ is given by the equation \cite%
{Gruner2000} $|U\chi (Q_{CDW1},T_{c})|=1$. Since the susceptibility increases
with the decrease of temperature, the largest value of $\chi$
corresponds to the highest value of CDW transtition temperature. We assume that the  electron-electron
interaction constant $U$ remains almost the same for considered series of 
RTe$_{3}$ compounds, because they have very close electronic structure. 
The result obtained (see Fig. \ref{FigChitPar2}) is comparable to the change of
transition temperature $T_{CDW1}$ observed in the RTe$_3$ series\cite{Ru2008}. 
The value $t_{\parallel}=1.85eV$ in $DyTe_{3}$ is the reference point. 
The experimentally observed transition temperature to $CDW_{1}$
state in TmTe$_{3}$ is $T_{CDW1}=245K$, while for GdTe$_{3}$ it is $T_{CDW1}=380K$
and for DyTe$_{3}$ it is $T_{CDW1}=302K$.\cite{Ru2008} This transition temperature is reduced by
35\% of its maximum value from GdTe$_{3}$ to TmTe$_{3}$. Thus, we may assume that
this range of $t_{\parallel}$ describes the whole series of compounds from TmTe$_{3}$
to GdTe$_{3}$. Moreover, basing on our calculations, we predict the values 
$t_{\parallel}\approx 1.37 eV$ in GdTe$_{3}$, $t_{\parallel}\approx 1.96 eV$ in HoTe$_{3}$, $%
t_{\parallel}\approx 2.06 eV$ in ErTe$_{3}$, and $t_{\parallel}\approx 2.20 eV$
in TmTe$_{3}$.

\begin{figure}[tbh]
	\centering
	\includegraphics[width=0.78\linewidth]{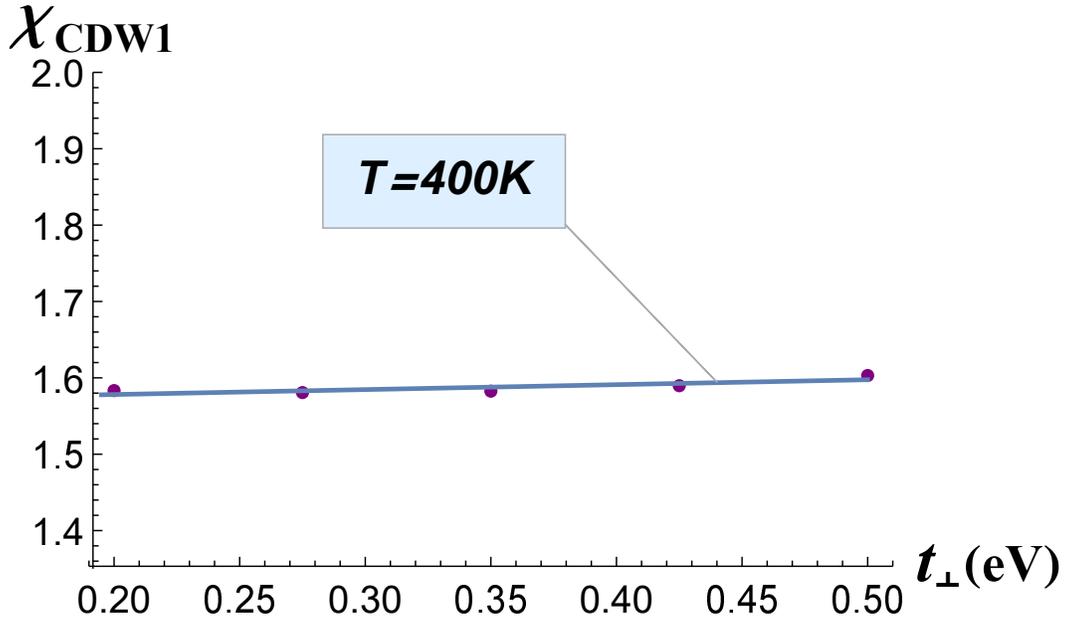} 
	\caption{Electron susceptibility $\protect\chi$ calculated at T=400 K
		as a function of the electron hopping term $t_{\perp}$.} 
	\label{FigChitPar3} 
\end{figure}

\begin{figure}[tbh]
	\centering
	\includegraphics[width=0.78\linewidth]{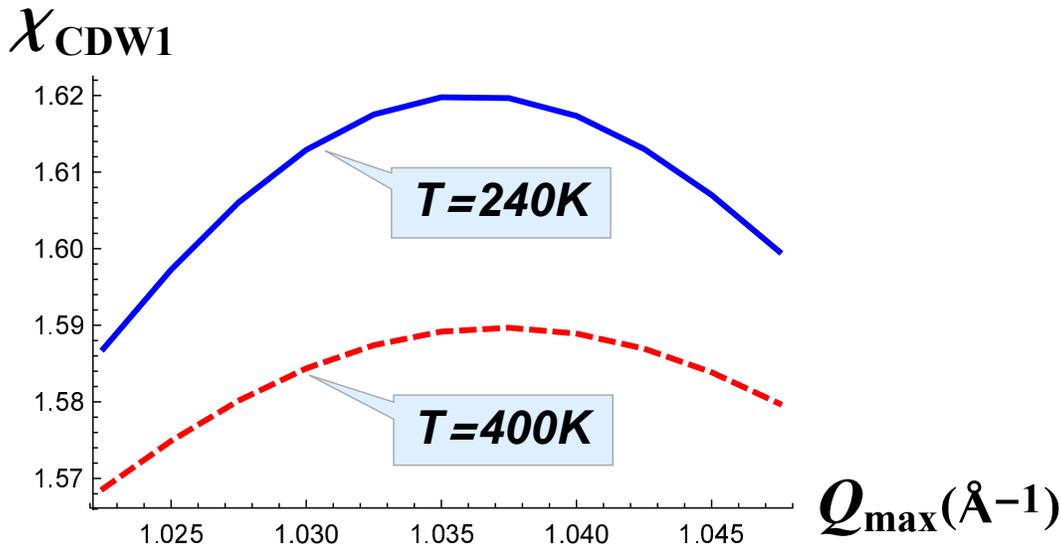} 
	\caption{The total susceptibility as a function of wave vector $Q_{max}$ near its maximum calculated at T=240 K (solid blue line) and at T=400 K (dashed red line).} 
	\label{QpT} 
\end{figure}

The dependence $\chi (t_{\perp})$ calculated at temperature above the transition is shown in Fig.\ref{FigChitPar3}. It is important to note there that with the decrease of temperature the wave vector does not shift and thus does not change its value, as it shown in Fig.\ref{QpT}: the position of maximum of susceptibility is almost the same for two different temperatures. Thus, the electronic susceptibility in Fig.\ref{FigChitPar3} is calculated on the same $Q_{max}$ wave vectors in Fig.\ref{FigChitPerp1}, but has lower value with the increase of temperature from 240 K to 400 K. 

The transition temperatures and conducting band parameters for varies RTe$_{3}$ compounds are summarized in Table.\ref{table}. The $t_{\parallel}$ increases with the increase of the atomic number of rare-earth element R; the $t_{\perp}$ decreases with the increase of the atomic number of rare-earth element R according to  $Q_{max}$($t_{\perp}$), but since the electronic susceptibility does not depend on $t_{\perp}$ we cannot calculate the values.

\begin{table}
	\centering
	\caption{List of parameters describing the dispersion and the CDW transition temperatures $T_{CDW2}$ and $T_{CDW2}$ for R = Gd, Dy, Ho, Er, Tm.}
	\label{table}
	\begin{tabular}{ | l | l | l | l | l | l | l |}
		\hline
		Compound & $T_{CDW1}$\cite{NRu2008}, K & $T_{CDW2}$\cite{NRu2008}, K & Lattice parameter \cite{NRu2008}, $\si{\angstrom}$  & $t_{\|}$, eV & $t_{\perp}$, eV & $E_{F}$, eV \\ \hline
		GdTe$_{3}$ & 377 & - & 4.320 & $\approx$1.37 & > 0.35 & 0.95 \\
		DyTe$_{3}$ & 306 & 49 & 4.302 & 1.85\cite{Brouet2008} & 0.35\cite{Brouet2008} & 1.28\\
		HoTe$_{3}$ & 284 & 126 & 4.290 & 1.96 & < 0.35 & 1.35\\
		ErTe$_{3}$ & 267 & 159 & 4.285 & 2.06 & < 0.35 & 1.42\\ 
		TmTe$_{3}$ & 244 & 186 & 4.275 & 2.20 & < 0.35 & 1.52\\ 
		\hline
	\end{tabular}
\end{table}

Our suggested values of the transfer integrals $t_{||}$ and $t_{\perp}$ assume that the effective electron-electron interaction at CDW wave vector does not depend considerably on the rare-earth element R.

\section{Conclusions}

To summarize, we have calculated the electron susceptibility on CDW$_1$ wave vector in the rare-earth tritelluride compounds as a function of temperature, wave vector, and two tight-binding parameters ($t_{\|}$ and $t_{\perp}$) of the electron dispersion. 
From these calculations we have shown that the parameter $t_{\perp}$ has almost no effect on the CDW$_1$ 
transition temperature $T_{CDW1}$ and weakly affects the CDW$_1$ wave vector $Q_{CDW1}$. On contrary, the variation of parameter $t_{\|}$ with the atomic number $n$ of rare-earth element drives the variation of both $T_{CDW1}$ and $Q_{CDW1}$. Note that the increase of $t_{\parallel}$ and of $t_{\perp}$ has opposite effects on $Q_{CDW1}$. Using the experimentally measured transition temperatures $T_{CDW1}$ , we estimated the values of $t_{\parallel}$ from our calculations for the whole series of RTe$_3$ compounds from TmTe$_{3}$ to GdTe$_{3}$.





\acknowledgements
The work was partially supported by the RFBR grants No. 19-02-01000, 17-52-150007 and 18-02-00280, by the Ministry of Education and Science of the Russian Federation in the framework of Increase Competitiveness Program of NUST ''MISiS'', and by the Foundation for Advancement of Theoretical Physics and Mathematics ''BASIS''.  P.D.G. thanks the State assignment 0033-2019-0001 ''The development of condensed-matter theory''. V.V.K. acknowledges Act 211 Government of the Russian Federation, contract No. 02.A03.21.0011.




\end{document}